\documentclass[12pt]{article}
\usepackage{a4}
\usepackage{amsmath}
\usepackage{amssymb}
\usepackage{amsfonts}
\begin{document}
\begin{center}
{\large\bf Euclidean formulation of relativistic quantum mechanics}\\[0.5cm]
{\bf W. N. Polyzou}$^1$ and Philip Kopp$^1$  \\[0.3cm]
$^1$Department of Physics and Astronomy,\\
Iowa City, IA 52246 \\[0.3cm]
\end{center}
We discuss preliminary work on a formulation of relativistic quantum
mechanics that uses reflection-positive Euclidean Green functions or
generating functionals as phenomenological input.  This work is
motivated by the Euclidean axioms of quantum field theory \cite{os}\cite{fr}.
The key observations are (1) locality is not used to
reconstruct the quantum theory and (2) it is possible to construct
a fully relativistic quantum theory without performing an explicit
analytic continuation.

Hilbert space vectors are represented by wave functionals $A[\phi]$
with inner product
\[
A[\phi] = \sum_{j=1}^{n_a} a_j e^{i \phi (f_j) }
\qquad
\langle A \vert B \rangle :=
\sum_{j,k}^{n_a,n_b}  a_l^* b_k Z[g_g-\Theta f_j]
\]
where $a_j$ are complex constants, $f_j$ are real Schwartz functions on 4
dimensional Euclidean space with positive-time support, $\Theta$ is
the Euclidean time-reflection operator, and $Z[f]$ is the Euclidean
generating functional.  Reflection positivity is the condition that
$\langle A \vert A \rangle \geq 0$.  For $\beta \geq 0$ and
$\mathbf{a} \in \mathbb{R}^3$ we define
\[
T(\beta, \mathbf{a}) A[\phi] := 
\sum_{j=1}^{n_a} a_j e^{i \phi (f_{j,\beta,\mathbf{a}}) }
\qquad 
f_{j,\beta,\mathbf{a}} (\tau, \mathbf{x}) := 
f_{j} (\tau-\beta , \mathbf{x}-\mathbf{a}) . 
\]
The square of the mass operator operating on a wave functional 
$A[\phi ]$ is 
\[
M^2 A[\phi] = \left ({\partial^2 \over \partial \beta^2} +
{\partial^2 \over \partial \mathbf{a}^2} \right )
T(\beta, \mathbf{a}) A[\phi]_{\vert \beta = \mathbf{a}=0}. 
\]
Solutions of the mass eigenvalue problem with eigenvalue $\lambda$ can
be expanded in terms of an orthonormal set of wave functionals
$A_n[\phi]$, $\langle A_n \vert A_m \rangle = \delta_{mn}$: 
\[
\Psi_{\lambda} [\phi] = \sum \alpha_n A_m [\phi] .
\]
Simultaneous eigenstates of mass, linear momentum, spin, and $z$ component of 
spin can be constructed from $\Psi_{\lambda} [\phi]$ using 
\[
\Psi_{\lambda, j, \mathbf{p},\mu }[\phi]=
\int_{SU(2)} dR \int_{\mathbb{R}^3}
{d\mathbf{a} \over (2 \pi )^{3/2}}  
e^{-i \mathbf{p} \cdot R \mathbf{a}}
U(R) T(0, \mathbf {a})\Psi_\lambda [\phi] 
D^{j*}_{\mu j}[R] 
\]
where $U(R)$ rotates the vector arguments of $f_j(\tau,\mathbf{x})$ in
$A[\phi]$.  When $\lambda$ is in the discrete spectrum of $M$,  
$\Psi_{\lambda, j, \mathbf{p},\mu }[\phi]$ is a wave functional for
a single-particle state that necessarily transforms as a mass 
$\lambda$ spin $j$ 
{\it irreducible representation}.

Products of suitably normalized single-particle wave functionals
define mappings from the product of single-particle irreducible
representation spaces of the Poincar\'e group to the model Hilbert
space.  Because these wave functionals create only single particle
states out of the vacuum, their products are Haag-Ruelle injection
operators \cite{jost}\cite{simon} for the two-Hilbert-space formulation 
\cite{simon} of scattering theory.
If we define $\Phi [\phi] := \prod_k \Psi_{\lambda_k, j_k,
\mathbf{p}_k,\mu_k }[\phi]$, $\otimes g_k = \prod g_k
(\mathbf{p}_k,\mu_k)$, and $H_f= \sum_k \sqrt{\lambda_k^2 +
\mathbf{p}_k^2}$, then scattering wave operator can be defined by the
limit
\[
\Omega_{\pm} \vert \otimes g_k  \rangle :=  
\lim_{t \to \pm\infty} e^{iHt} \Phi e^{-iH_ft} \vert \otimes g_k \rangle.
\]
Using the Kato-Birman invariance principle \cite{simon} to replace 
$H$ by  $-e^{-\beta H}$ gives
\[
\Omega_{\pm} \vert \otimes g_k \rangle :=  
\lim_{n \to \pm\infty} e^{-ine^{-\beta H}}
\Phi e^{ine^{-\beta H_f}} \vert \otimes g_k \rangle . 
\]
Since the spectrum of $e^{-\beta H}$ is compact, for large {\it fixed} $n$ 
$e^{-ine^{-\beta H}}$ can be uniformly approximated by a polynomial in 
$e^{-\beta H}$, which is easy to calculate in this framework.
These steps provide a means to construct all single-particle states,
all scattering states, and compute the action of the Poincar\'e group
on all single-particle states and $S$-matrix elements, using only
the Euclidean generating functional as input.
The advantages of this framework are the relative ease with which cluster 
properties can be satisfied, the close relation to the quantum mechanical 
interpretation of quantum field theory,  and the ability to perform 
calculations directly in Euclidean space without analytic continuation. 

We tested the general method for calculating scattering observables
using a solvable quantum mechanical model of the two-nucleon system.
These test calculations, which used narrow wave packets, a large
finite $n$ and a Chebyshev polynomial expansion of $e^{inx}$,
exhibited convergence to the exact transition matrix elements for a range
of relative momenta between about 100 MeV up to 2 GeV.  This success
warrants further investigation of this framework.

This research was supported by the U.S. Department of Energy.

\end{document}